\documentclass[%
 reprint,
 amsmath,amssymb,
 aps,
pra,
]{revtex4-2}

\usepackage{graphicx}
\usepackage{dcolumn}
\usepackage{bm}
\usepackage{caption}
\usepackage{subcaption}
\usepackage{multirow}
\usepackage{adjustbox}
\usepackage{ragged2e}
\usepackage{xcolor}
\usepackage{float}

\begin{document}

\preprint{APS/123-QED}

\title{
A comparison of Fraunhofer-type diffraction from an atomic single-slit and a molecular double-slit}.

\author{Jibak Mukherjee$^{1}$, Kamal Kumar$^{1}$, Harpreet Singh$^{1}$, Manojit Das$^{1}$, Guo-Peng Zhao$^{2}$, Ling Liu$^{3}$, Károly Tőkési$^{4}$, Lokesh C. Tribedi$^{1,5}$, and Deepankar Misra$^{1}$ }
 \email{dimsra@tifr.res.in}
 
\affiliation{
 $^{1}$Department of Nuclear and Atomic Physics, Tata Institute of Fundamental Research, Mumbai 400005, India.\\
 $^{2}$College of Data Science, Jiaxing University, Jiaxing 314001, China\\
 $^{3}$Data Center for High Energy Density Physics, Institute of Applied Physics and Computational Mathematics, Beijing 100088, China\\
 $^{4}$HUN-REN Institute for Nuclear Research (ATOMKI), 4026 Debrecen Bemtér 18/c, Hungary\\
 $^{5}$CIDRI, R\&D, UPES, Bidholi campus, Dehradun, 248007, Uttarakhand, India}

%




\date{\today}

\begin{abstract}

We measured the Q-value and the scattering angle distributions for non-dissociative state selective single electron capture in collisions of 7.5 keV $\mathrm{H^+}$ and 15 keV $\mathrm{H_2^+}$ with He. The experimental data are compared with semiclassical close-coupling calculations and predictions from the classical trajectory Monte Carlo simulations. By analogy with Fraunhofer diffraction, we also developed a toy model to reconstruct an imaginary screen that reflects the reaction impact-parameter dependence, in channels where the magnetic quantum number remains unchanged. It is well established that $\mathrm{H_2^+}$ acts as a molecular double-slit in scattering processes.  By demodulating the Young's double-slit-type interference pattern, we extracted the individual slit diffraction pattern of $\mathrm{H_2^+}$ and compared it with that of the $\mathrm{H^+}$ atomic single-slit. For ground state electron capture, we found that the single and the double slit diffraction patterns have equal fringe width, whereas for excited state electron capture, diffraction patterns are quite different. 
\end{abstract}

\maketitle

 When two identical atoms associate to form a homonuclear diatomic molecule, the electron orbitals possess an inversion symmetry about the center of symmetry point. Within the framework of the Linear Combination of Atomic Orbitals (LCAO) approximation, the molecular wave function is constructed from the atomic orbitals of the constituent atoms as $\mathrm{\Psi_{molecule} = (\Psi_a \pm \Psi_b})$ where $\mathrm{\Psi_a}$ and $\mathrm{\Psi_b}$ are atomic orbitals centered around two nuclei a and b of the diatomic molecule, the $\pm$ sign stands for gerade and ungerade states. This inherent symmetry manifests directly in scattering processes as a two-center interference effect, the atomic orbitals $\mathrm{\Psi_a}$ and $\mathrm{\Psi_b}$ act as coherent, identical scattering sources separated by the internuclear distance $\mathrm{\boldsymbol{\rho}}$. The molecular T-matrix in this case is written as \cite{book3}
\begin{equation}\label{eqn:T_molecule}
    T_{molecule} = \left(1 \pm e^{\boldsymbol{i\Delta K}\cdot \boldsymbol{\rho}}\right) T_{atomic},
\end{equation}
where $\mathrm{T_{atomic}}$ is the identical T-matrix for each scattering center, $\boldsymbol{\Delta K}$ is the change in the wave vector in the scattering processes.  In a seminal work by Cohen and Fano \cite{PhysRev.150.30}, they first established the idea of Young's double-slit-type interference effect in photoionization processes of diatomic molecules. This foundational prediction of two-center interference has since been validated across a diverse range of studies, including photo ionization\cite{doi:10.1126/science.1144959, doi:10.1073/pnas.1018534108,PhysRevLett.103.043001}, ion-molecule collision\cite{PhysRevLett.87.023201,PhysRevLett.92.153201,A_B_Wittkower_1966} and electron impact molecule ionization \cite{PhysRevA.78.052701,PhysRevA.80.062704}.

The occurrence of two-center interference extends to state-selective electron capture processes where molecules are involved. Tuan and Gerjuoy ~\cite{PhysRev.117.756} first predicted it in $\mathrm{H^+ + H_2}$ collision system. Later, it was substantiated by many theoretical \cite{PhysRevA.40.1302,PhysRevA.38.3769,PhysRevA.23.1807,PhysRevA.40.3673} and experimental \cite{PhysRevA.47.3923,PhysRevA.72.050703,PhysRevA.96.042708,D_Fischer_2007,PhysRevLett.101.083201,PhysRevLett.102.153201} studies with atomic projectile and molecular target systems. Notably, this effect also emerges distinctly in the inverse scenario involving molecular projectile and an atomic target~\cite{PhysRevLett.101.173202,PhysRevLett.112.023201}. 

Schmidt and coworkers observed  Young's double-slit-type interference pattern in the transverse momentum plane of 10 keV $\mathrm{H_2^+ + He}$  collision system \cite{PhysRevLett.101.173202}. Dissociation following $\mathrm{H_2(b^3\Sigma^+_u)}$ state selective electron capture made it possible to map the interference pattern in the transverse momentum plane as a function of molecular orientation.  It was found out that some minima in the 2D transverse momentum distribution do not arise from the destructive interference effect, but rather from the single-center diffraction pattern that is also observed in the $\mathrm{H^+ + He}$ charge exchange collision.  Johnson \emph{et al.} ~\cite{PhysRevA.40.3626} measured the absolute differential cross section of charge exchange in the 5 keV H$^+ + $He collision. In their study, oscillatory scattering angle distribution, and from theoretical analysis, they concluded that the dominant contribution to the absolute differential cross section comes from electron capture to the H (n = 1) state.

However, a significant question remains unanswered in the literature: namely, how does the diffraction pattern change in state-selective electron capture processes when a molecular ion captures an electron, compared to the pattern from its constituent atomic ions? Stated more precisely, it is unclear whether the transition matrix element, $\mathrm{T_{atomic}}$ in Eq.~(\ref{eqn:T_molecule}), can be accurately modeled using the T-matrix of an isolated atomic ion. To investigate this, we have performed collision experiments between 7.5 keV/u $\mathrm{H^+}$ and $\mathrm{H_2^+}$ projectiles and the He target atom. Within this energy range, the impact parameter model with the Eikonal approximation is legitimately applicable, where the nuclear motions  can be approximated by a set of rectilinear trajectories defined by:~ \cite{PhysRevLett.87.123201,book1,book2,R_McCarroll_1968,PhysRev.169.84},
\begin{align}\label{eqn:Eikonal_approx}
    \boldsymbol{R} &= \boldsymbol{b} + \boldsymbol{v_0}t  ~~ \mathrm {and}\\
     \boldsymbol{v_0}\cdot\boldsymbol{b} &= 0,
\end{align}
where $\boldsymbol{R}$ is the internuclear separation between the target and the projectile at the time instance t, $\boldsymbol{v_0}$ is the relative velocity between them, and $\boldsymbol{b}$ is the impact parameter. The equal relative velocity between the target and the projectile gives an identical time dependence of the perturbative potential. Therefore, by comparing the resulting diffraction patterns of $\mathrm{H}$ and $\mathrm{H_2}$, we can directly probe the alteration of the matter wave during the scattering process. In this context, any deviation from the rectilinear motion is defined as matter wave diffraction.

In this work, we focus solely on the non-dissociative, state-selective electron capture events. As a consequence, the measured scattering angle distribution in $\mathrm{H_2^+ + He}$ collision will be an average distribution, and the fringe visibility of the orientation-averaged interference pattern will be less pronounced compared to the fixed orientation interference pattern. The differential cross sections have the form~\cite{book3,PhysRev.150.30}
\begin{align}
    \left<\dfrac{d\sigma_{i\rightarrow f}}{d\Omega}\right>_{molecule} &= 2I_{i\rightarrow f}(\Omega)\left<\dfrac{d\sigma_{i\rightarrow f}} {d\Omega}\right>_{atomic} \mathrm {and} \label{eqn:dndo_molecule} \\
    I_{i\rightarrow f}(\Omega) &= 1 + \dfrac{sin(\Delta K \rho)}{\Delta K \rho}, \label{eqn:intf}
\end{align}

where the identical center diffraction pattern $\left<\frac{d\sigma_{i\rightarrow f}}{d\Omega}\right>_{atomic}$ is modulated by the interference pattern $I(\Omega)$, and $\Delta K = \sqrt{\Delta k_\parallel^2 + \Delta k_\perp^2}$.

 
Our present experiments were carried out in the electron cyclotron resonance-based ion accelerator (ECRIA) facility~\cite{Agnihotri_2011} at the Tata Institute of Fundamental Research, Mumbai, using the home-built Cold Target Recoil Ion Momentum Spectroscopy (COLTRIMS) setup~\cite {10.1063/5.0100395}. Briefly, $\mathrm{H^+}$ and $\mathrm{H_2^+}$ ions were produced in a 14.5 GHz electron cyclotron resonance (ECR) ion source. The ions were extracted at extraction voltages of 7.5 kV and 15 kV for $\mathrm{H^+}$ and $\mathrm{H_2^+}$, respectively, mass analyzed using a $\mathrm{90^0}$ dipole magnet and directed towards the collision chamber. Before entering the collision chamber, the ion beam was subjected to additional purification using a $30^0$ cylindrical electrostatic charge state analyzer. Inside the collision chamber, the projectile ion beam collided with the cold target in the cross-beam configuration. The target was prepared using He gas supersonic expansion through a 30 $\mathrm{\mu m}$ nozzle and passed through 2 skimmers for geometric cooling purposes. After the reaction, the recoil ions were extracted perpendicular to both the projectile beam and the supersonic molecular jet, using a 5.33 V/cm electric field and further maneuvered toward the recoil ion time and position sensitive detector  $\lbrace$MCP+HEX from Roentdek$\rbrace$ such that momentum components have one-to-one mapping with the time of flight (tof) and the detector hit position data. After the collision, the non-interacting projectile beam was separated using an electrostatic charge state deflector, and the neutral projectile, after electron capture, hit the time and position sensitive projectile detector  $\lbrace$MCP+DLD by Roentdek$\rbrace$, located approximately 1 meter downstream of the collision region. The recoil ion momentum was measured in coincidence with the projectile. The estimated momentum resolutions were $\delta p_x, \delta p_z \sim $0.2 a.u. and $\delta p_y \sim $1.6 a.u., where x is the projectile ion direction, y is the supersonic jet direction, and z is along the spectrometer axis. From the $p_x$ momentum component, we deduced the Q-value by the relation
\begin{equation}
    Q = -v_p p_x - \dfrac{v_p^{2}}{2}.
\end{equation}
The transverse momentum ($p_\perp$) distribution was derived from the projection $p_z$ using the inverse Abel transformation ~\cite{Alessi_2015}. The projectile scattering angle in the laboratory frame was derived from $p_\perp$ using the relation
\begin{equation}
    \theta = \dfrac{p_\perp}{m_p v_p},
    \label{eqn:theta}
\end{equation}
where $m_p$ and $v_p$ are the projectile mass and velocity in the laboratory frame.


In  $\mathrm{H_2^+ + He}$ collision, there is a possibility of the dissociation of the neutral $\mathrm{H_2}$ molecule into two H fragments following one electron capture \cite{Vogler1978,PhysRevLett.101.173202}. However, in the present work, we will only consider the non-dissociative single-electron capture channels. The reactions can be written as follows

\begin{align}
H^+ + He(1s^2) &\rightarrow H(n_f) + He^+(1s) \\
    H_2^+(n_i,\Lambda_i,\upsilon_i) + He(1s^2) &\rightarrow H_2(n_f,\Lambda_f,\upsilon_f) + He^+(1s)
    \end{align}

where $n_i$ and $n_f$ are the initial and final principal quantum numbers of the Hydrogen atom and molecule, $\upsilon_i, \upsilon_f$ are the vibrational quantum numbers, $\Lambda_i$ and $\Lambda_f$ are the magnetic quantum numbers of $
\mathrm{H_2^+}$ and $\mathrm{H_2}$, respectively.

\begin{figure}
    \centering
    \includegraphics[width=\linewidth]{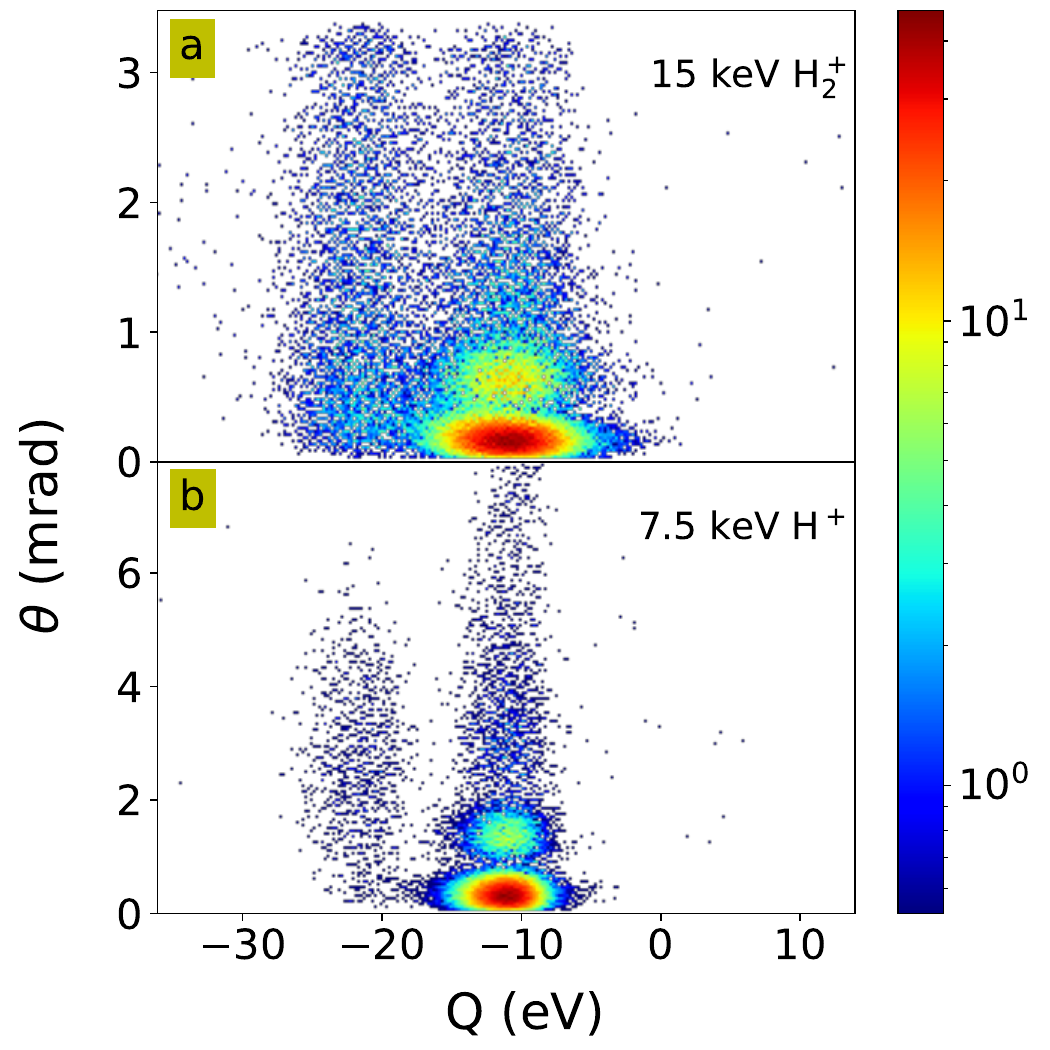}
\caption{\justifying{Density plot of Q-value with scattering angle for 7.5 keV/u projectiles. Panel (a) presents spectrum for $\mathrm{H_2^+}$ and (b) presents spectrum for $\mathrm{H^+}$.}}
\label{fig:Q_He}
\end{figure}

The measured counts, differential in two observables, the Q-value, and the scattering angle $\mathrm{\theta}$, are plotted in Fig.(\ref{fig:Q_He}). We can notice traces of two resolved channels along the Q-value axis for both $\mathrm{H^+}$ and $\mathrm{H_2^+}$ projectiles. In $\mathrm{H^+ + He}$, the broadening in the Q value is mainly due to the resolution of the spectrometer. It is more in $\mathrm{H_2^+ + He}$ because the closely lying vibrational energy levels are in an unresolved condition in the present experiment.

The Q-value of the inelastic collision is defined as the total binding energy difference between the initial and final systems. In the $\mathrm{H^+ + He}$ collision system, the highest populated state with Q = -11 eV is the ground state of H($n_f = 1$). The lower populated state around Q = -21.2 eV is the first excited state of H($n_f = 2$). At projectile velocity $v_p=0.548$ a.u., our spectrometer resolution was $\mathrm{\Delta Q\sim 3\ eV}$. So, there may be upper excited states ($n_f = 2,3,...$) in an unresolved condition. But their partial cross sections are less than 1$\%$~\cite{H_A_Slim_1991}.

In $\mathrm{H_2^+ + He}$ collision, the highest populated state with Q $\simeq$ -10.5 eV is the electronic ground state of $\mathrm{H_2 (X ^1\Sigma_g^+ 1s\sigma^2)}$, while the lowest populated state with Q $\simeq$ -21.5 eV can be any of the excited states among $\mathrm{H_2((E ^1\Sigma_g^+ 2s\sigma) / (F ^1\Sigma_g^+ 2p\sigma^2)/(C ^1\Pi_u 2p\pi))}$. From the Q-values, it is also evident that before collision, $\mathrm{H_2^+}$ was in the electronic ground state $\mathrm{X ^2\Sigma_g^+ 1s\sigma}$.

\begin{figure*}
    \centering
    \includegraphics[width=0.5\textwidth]{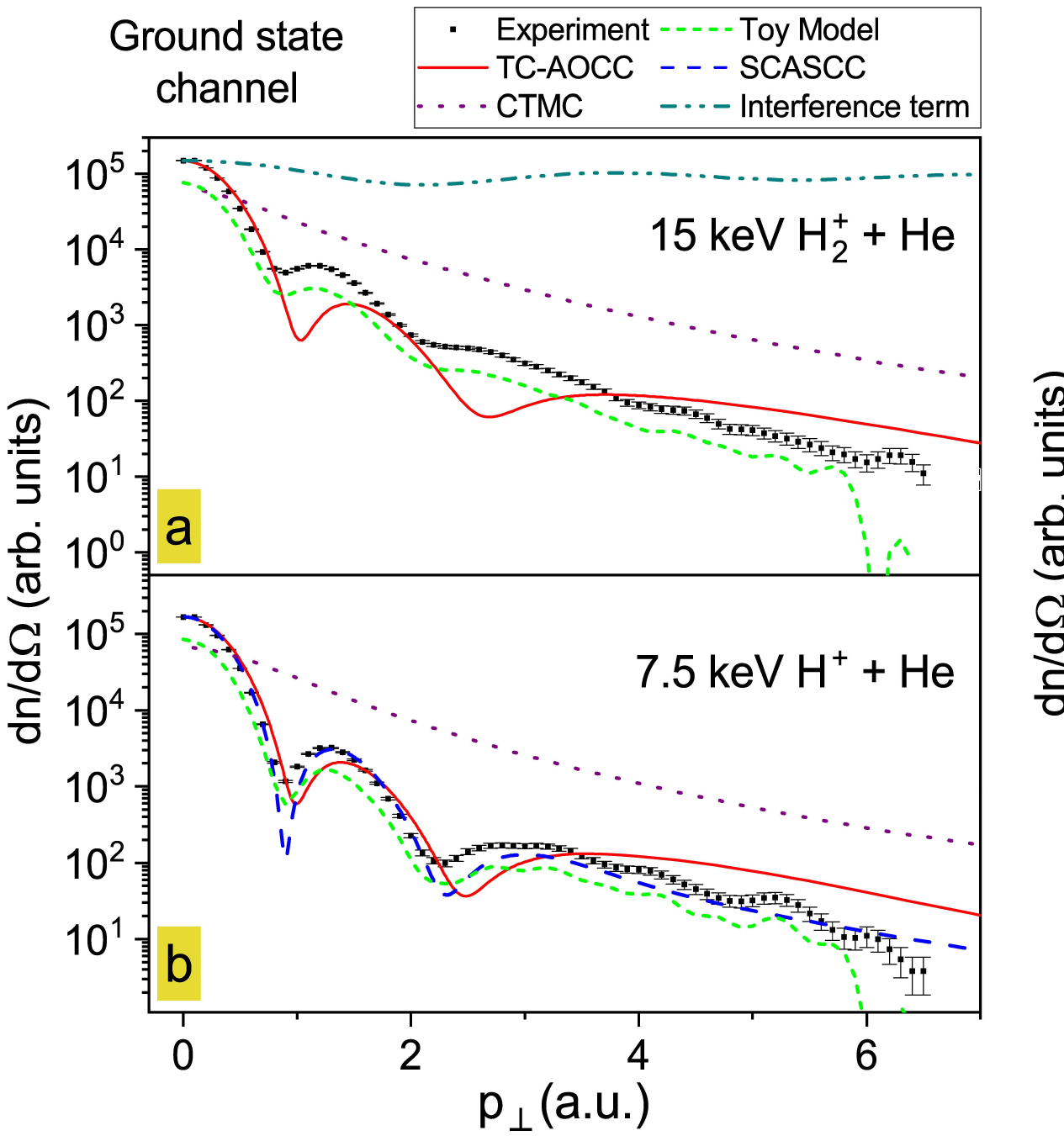}
\caption{\justifying{Measured state selective differential cross-sections along with theoretical predictions. Panels (a) and (b)  present results for electron capture to the ground state of $\mathrm{H_2}$ and H, respectively. Panels (c), (d) present the results for electron capture to the excited electronic states of the same projectiles. The experimental data are represented in black solid square, the TC-AOCC results in a solid red curve, the SCASCC results in a blue dashed curve,  CTMC simulation results in purple dotted curves. Results derived from our toy model for the ground state electron capture are plotted in panels (a) and (b) in a light green short dashed curve with a little offset for better visibility. The orientation averaged state selective double-slit interference patterns for $\mathrm{H_2^+}$ projectile are plotted in dark cyan dash-double-dotted curves for inter-nuclear separation of 2 a.u., in panels (a) and (c). Error bars in the experimental data are statistical in nature ($\sqrt{n}$).}}
\label{fig: scattering all}
\end{figure*}

Hereafter, we refer to the capture of an electron by $\mathrm{H^+}$ and $\mathrm{H_2^+}$, to the ground electronic state as the "ground state channel" and to the excited electronic states as the "excited state channel". Fig. (\ref{fig: scattering all}) shows the measured transverse momentum distributions along with theoretical predictions. Unless specified, we have used the atomic units (a.u.) ($\hbar = m_e = e^2= 1$). 

 The diffraction patterns from the $\mathrm{H^+}$ projectile is directly reflected in the structures of state selective differential cross section.  Fig.~\ref{fig: scattering all}(b),~\ref{fig: scattering all}(d) depict the measured differential cross section along with theoretical predictions. The structures can be explained in terms of the integral representation of the differential cross section in the impact parameter model with the Eikonal approximation ~\cite{R_McCarroll_1968,PhysRev.169.84,book1,book2}.  For a cylindrically symmetric interaction about the projectile-target relative velocity axis, the scattering amplitude in the center of mass frame can be written as 
\begin{multline}\label{eqn:ftp}
    f_{i\rightarrow f}(\Theta,\Phi) = -iK exp [ -i\Delta m \Phi ] (i)^{\Delta m}  \\
    \times \int_0^\infty b db J_{\Delta m}(K\Theta b)\tilde{\mathcal{A}}_{i\rightarrow f}(b)exp[2i\delta_{i\rightarrow f}(b)] 
\end{multline}
\begin{equation}{\label{eqn:dndt}}
    \dfrac{d\sigma_{i\rightarrow f}}{d\Omega}(\Theta,\Phi) = \left|f_{i\rightarrow f}(\Theta,\Phi)\right|^2,
\end{equation}
where $\Theta$ is the scattering angle, $\Phi$ is the azimuthal angle, $v_0$ is the relative velocity between the projectile and the target, $\mu$ is the reduced mass of the target projectile combination,  $K = \mu v_0$, b is the impact parameter, $\delta_{i\rightarrow f} (b)$ is the Eikonal phase change, $\tilde{\mathcal{A}}_{i\rightarrow f}(b)$ is the probability amplitude. In the reaction if the electron was in the initial state with magnetic quantum number $m_i$ and transferred to a state with magnetic quantum number $m_f$, then the dependence of $\Delta m = m_i - m_f$ appears in the azimuthal phase $e^{-i\Delta m \Phi}$ and as the $\Delta m$$^{th}$ order Bessel function of the first kind $J_{\Delta m}$.

In the $\mathrm{H^+}$ ground state channel, electron transfer occurs with $\Delta m = 0$ ($\mathrm{He (1s^2)\rightarrow H (n=1)}$). For this, the diffraction pattern (Fig.~\ref{fig: scattering all} (b)) has a line shape like $\left|\int \tilde{\mathcal{A}_0}(b)J_0(b)db\right|^2$. This is analogous to Fraunhofer diffraction from a circular aperture ~\cite{PhysRevLett.87.123201,M_van_der_Poel_2002}.
 In the $\mathrm{H^+}$ excited state channel, $\Delta m = 0, \pm 1$ are possible ($\mathrm{He (1s^2)\rightarrow H (n=2,l=0,1)}$). Consequently, the diffraction pattern (Fig.~\ref{fig: scattering all} (d)) exhibited shape like $\left|\int \tilde{\mathcal{A}_0}(b)J_0(b)db\right|^2 + \left|\int \tilde{\mathcal{A}_1}(b)J_1(b)db\right|^2$.  These types of pattern have also been observed in other channel-resolved differential cross-section measurements, for example, in collision systems such as $\mathrm{C^{4+} + He}$ ~\cite{PhysRevA.103.032827}, $\mathrm{O^{4+} + He}$ ~\cite{PhysRevA.109.052811}, and $\mathrm{Ar^{8+} + He}$~\cite{PhysRevA.110.012809}.

The $\mathrm{H_2^+}$ projectile differential cross section consists of an orientation averaged double slit interference pattern ~\cite{PhysRev.117.756,PhysRevLett.101.173202} and the slit diffraction pattern, as described in Eq.  (\ref{eqn:dndo_molecule}) and Eq. (\ref{eqn:intf}). The calculated interference patterns for internuclear separation $\mathrm{\boldsymbol{\rho}=2\ a.u.}$ in the ground and excited state channels are plotted in Fig.~\ref{fig: scattering all} (a), \ref{fig: scattering all}(c), along with the measured differential cross sections. For $\nu_i = 0$  the most probable internuclear separation is $\rho = 2$ a.u.  ~\cite{PhysRevLett.108.073202}. It can be noticed that the fringe visibility of the interference pattern is much weaker compared to the structures of the differential cross section.  This indicates that the diffraction pattern is largely unmodulated by the interference pattern and it has been dominantly manifested in the measured differential cross section structures. Hence by analyzing the structures we can analyze the individual slit diffraction patterns.

The $\mathrm{H_2^+}$  ground state channel diffraction pattern resembles the structure of Fraunhofer type diffraction from a circular aperture. Since the $\mathrm{H_2(X ^1\Sigma_g^+ 1s\sigma^2)}$ state has $\Lambda_f=0$, the line shape like $\left|\int \tilde{\mathcal{A}_0}(b)J_0(b)db\right|^2$ is justifiable. However, in the $\mathrm{H_2^+}$ excited state channel, non-zero $\Lambda_f$ is defined along the molecular axis, not in the direction of $\boldsymbol{v_0}$. This implies that azimuthal phase dependence along $\boldsymbol{v_0}$ cannot be characterized by $\Lambda_f$, and Eq.~\ref{eqn:ftp} cannot be applied directly to evaluate the single-center scattering amplitude. 

An exact theoretical solution is not feasible; we have drawn quantum and classical pictures of the reaction dynamics by making approximations. In the quantum picture, we have used semiclassical close coupling approach in two different interaction scenarios. Foundational details are provided elsewhere \cite{Liu_2024,PhysRevA.110.032806}. Briefly, nuclear motions are approximated as straight line trajectories and active electrons' motion is described by the $\mathrm{Schr\ddot{o}dinger}$ equation. In the first scenario, Two Center Atomic Orbit Close Coupling (TC-AOCC) method, the active electron wave function is solved under the influence of projectile and target potentials. In the second scenario, the Semiclassical Asymptotic-State-Close-Coupling (SCASCC) method, taking care of electron-electron interaction, the two electron wave function is solved under target and projectile potential. Mathematical formulation is given in Appendix I.  
\paragraph*{} Foundational details of the Classical Trajectory Monte Carlo method is provided somewhere else \cite{R_Abrines_1966,PhysRevA.16.531,TOKESI1994201,K_Tökési_2000}. Briefly,
the CTMC calculations were
performed using three-body approximations. The many-electron target atom was simplified to a one-electron atom,
treating the projectile ion as a single particle. The Coulomb force acted among the colliding particles. 
Trajectories are initiated randomly, and classical equation motions were solved using the Runge-Kutta method. Further details are provided in appendix II.

\begin{figure}[htb]
    \centering   \includegraphics[width=\linewidth]{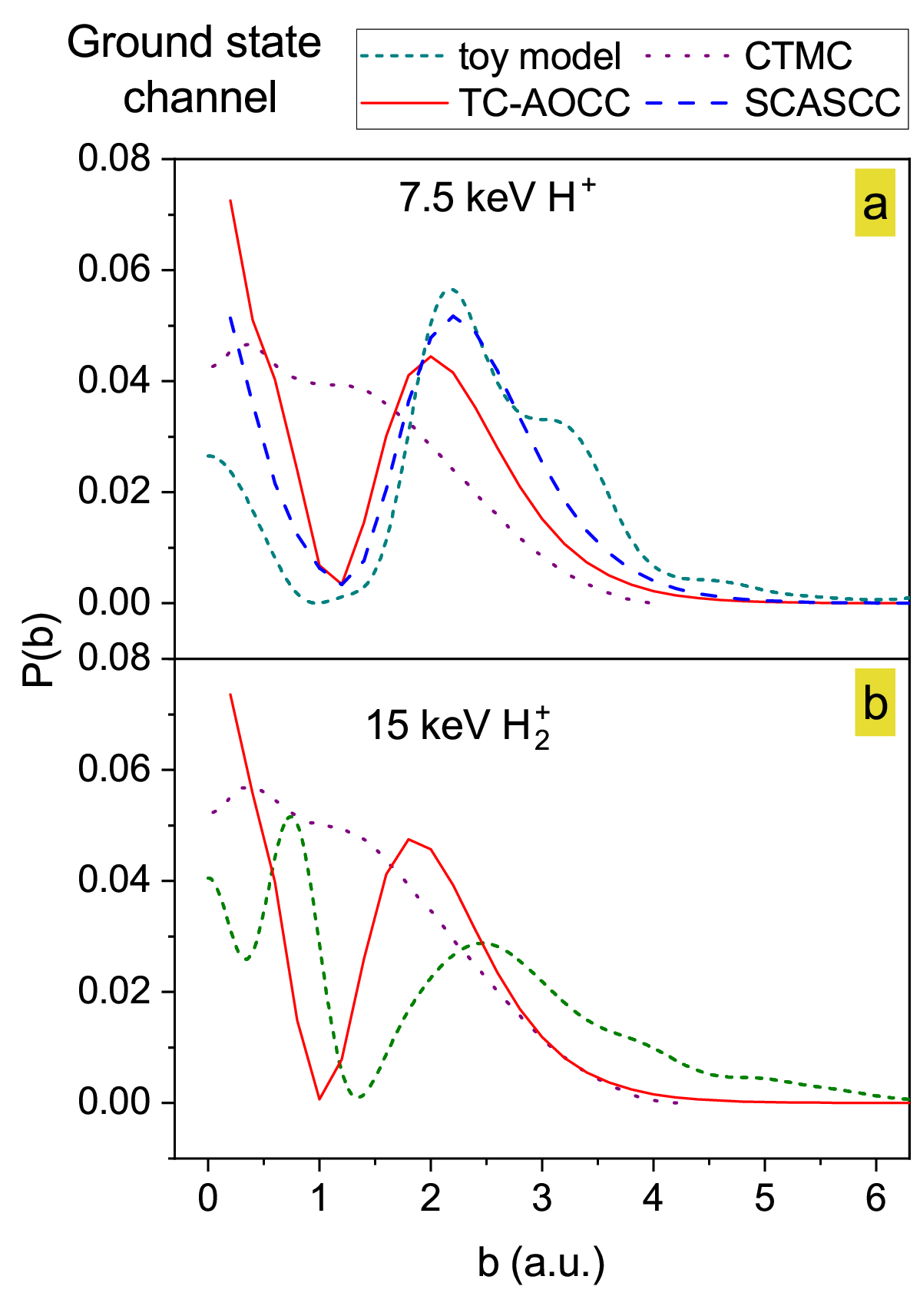}
    \caption{\justifying{Impact parameter dependent probability of electron capture to the ground state. Panel (a) presents results for 7.5 keV H$^+$ and (b) presents 15 keV H$_2^+$ results. The TC-AOCC, SCASCC, and CTMC results are represented by red solid, blue dashed, and purple dotted curves, respectively. The toy model estimations are represented by a green short-dashed curve. }}
    \label{fig:theory_gs}
\end{figure}

We have also developed a toy model for the ground state channel that assumes the incoming matter wave, having a plane wavefront passes through an imaginary Fraunhofer screen~\cite{PhysRev.169.84} which alters the amplitude and phase of the wavefront. Coherent superposition of the altered wavefront makes Fraunhofer diffraction from a circular aperture-like pattern~\cite{PhysRevLett.87.123201,M_van_der_Poel_2002}. Previously developed models ~\cite{PhysRevResearch.6.013108,Wang_2012} assumed the aperture function to be a constant pupil kind. We reconstructed an imaginary complex screen where individual slit follows Eq. (\ref{eqn:ftp}) ($\mathrm{H^+}$ has single slit and $\mathrm{H_2^+}$ has double slit). We computed the inverse Hankel transform using inverse curve fitting to the experimental data. The screen also reflects the impact parameter dependence. Details are given in Appendix III.  

The agreement of the theoretical calculations with the experimental data are depicted in Fig. (\ref{fig: scattering all}). We can see that the $\mathrm{H^+}$ ground state channel has a good agreement with TC-AOCC result and an excellent agreement with SCASCC result. This may be due to the fact that the Eikonal approximation works very well at low scattering angles and in the intermediate velocity region. Only the active electron motion in a central potential may not be a complete description; incorporation of electron-electron interaction term gives more accurate result. However, in the $\mathrm{H_2^+}$ ground state channel, the TC-AOCC method models the interaction of two scattering centers by an effective central potential, which may not be sufficient to get the accurate diffraction fringe widths. The classical description does not include a coherent superposition of matter waves. So the obtained differential cross section structures are the envelopes of the momentum space wave functions.  We can see that the $\mathrm{H^+}$ and $\mathrm{H_2^+}$ diffraction envelopes are well described by CTMC calculations. The toy model gives identical structures because it is derived from inverse curve fitting.

The impact parameter dependent probability distributions for the ground state channel are depicted in Fig. (\ref{fig:theory_gs}). The equal fringe widths of the diffraction pattern, in our measurement resolution, for $\mathrm{H^+}$ and $\mathrm{H_2^+}$ projectiles, resulting from a common feature at  $\mathrm{b > 1}$ a.u. in the probability distributions. Below 1 a.u., the impact parameter dependence is slightly different, indicating a different charge distribution closer to the nucleus. The range of probability distribution also suggests that the slit width of the individual slits of $\mathrm{H_2^+}$ is larger than the slit separation.

In our work, we aimed to explain the molecular diffraction pattern in an innovative way. We first analyzed the individual slit diffraction pattern and then reconstructed the double-slit diffraction pattern by incorporating the interference effect. The individual slits of $\mathrm{H_2^+}$ are nearly identical to those of $\mathrm{H^+}$ in the ground state channel at higher impact parameters, but differ at lower impact parameters. This suggests that modeling the individual molecular double-slit is feasible using the constituent atomic single-slit in the ground state channel. However, in the excited state channel, although the impact parameter-dependent probability distribution of electron capture may be nearly identical, the differing azimuthal phase dependence prevents us from modeling the molecular double-slit with an atomic single-slit. This finding will influence future modeling of molecular scattering. Additionally, it is intriguing to consider how matter waves are altered during charge exchange reactions, as this knowledge may be beneficial for future applications in wavefront shaping of matter waves.

In conclusion, we have measured the relative differential cross sections of resolved, non-dissociative, state-selective electron capture channels, which also reflect the diffraction patterns in equal velocity $\mathrm{H^+ + He}$ and $\mathrm{H_2^+ + He}$ collisions.  The much lower visibility of the orientation-averaged interference fringes enables us to directly access the diffraction patterns from the individual slits of $\mathrm{H_2^+}$. We found that for ground state electron capture, the diffraction pattern exhibited structures with equal fringe widths. Theoretically, we investigated the problem using a semiclassical close coupling approach and classical trajectory Monte Carlo method by approximating a central potential of $\mathrm{H_2^+}$, whereas in the toy model, we back-calculated single- and double-slits. The impact parameter dependence for the ground state channel showed nearly identical probability distributions above a certain value. In contrast, the excited state electron capture diffraction patterns are quite different, which may be due to different azimuthal phase dependence.

\emph{Acknowledgment} - The authors thank ECRIA lab members for technical support. This work is supported by the Department of Atomic Energy (Government of India) under Research Project No. RTI 4002.

\emph{Data Availability} - The data that support the findings of this study are available from the authors on reasonable request.


\bibliography{apssamp}

\section*{Appendix I}
\subsection*{TC-AOCC method}
In two center atomic orbit close coupling approach, nuclear motion is approximated by Eqn. \ref{eqn:Eikonal_approx}. The active electron wave function is described by 
\begin{equation} \label{eqn:TCAOCC_sch_eqn}
    \left[H_e - i\dfrac{\partial}{\partial t}\phi_e (\boldsymbol{r},t)\right]=0
\end{equation}
Where the electronic Hamiltonian is
\begin{equation}
\label{eqn:HAM_TCAOCC}
    H_e = -\dfrac{1}{2}\nabla^2 + V_A(\vec{r_A}) + V_B(\vec{r_B})
\end{equation}

and $V_{A}(\boldsymbol{r_A})$ and $V_{B}(\boldsymbol{r_B})$ represent the interactions between the active electron and the projectile($H^+$, $H_2^+$) and target(He), respectively. In the present calculation, these potentials are expressed as
\begin{align}
    V_{He} (r)&=-\dfrac{1}{r}-\dfrac{1}{r} e^{-2.69697r}-0.65354 e^{-2.69697r}\\
    V_{H_2^+}(r) &= -\dfrac{1}{r} -\dfrac{1}{r} e^{-6.4714 r}
\end{align}
For the H$^+$ projectile, the Coulomb potential has been used to describe this interaction. 
$\phi_e$ is expanded in atomic orbitals of moving projectile and target. 

\subsection*{SCASCC method}
In this method, the electronic equation of motion is 
\begin{equation}\label{eqn:eom_SCASCC}
    \left[H_e-i\dfrac{\partial}{\partial t}\right]\Phi_e(\boldsymbol{r}_1,\boldsymbol{r}_2,t) = 0
\end{equation}
where the Hamiltonian is
\begin{equation}
\label{Eq:ham_SCASCC}
    H_e = \sum_{i=1,2}-\dfrac{1}{2}\nabla_i^2+V_B(\boldsymbol{r}_l^l) + V_A(\boldsymbol{r}_l^P)+\dfrac{1}{\boldsymbol{r}_1-\boldsymbol{r}_2}
\end{equation}\label{eqn:SCASCC He}
where $\boldsymbol{r}_l$ and $\boldsymbol{r}_l^P = \boldsymbol{r}_l - \boldsymbol{R}(t)$ are the position vectors of the electrons with respect to the target and the projectile, respectively.

where $\Phi_e(\boldsymbol{r}_1,\boldsymbol{r}_2,t)$ is the total wave function of the collision system, and can be expanded in terms of a set of electronic states of isolated collision partners as

\begin{multline}\label{eqn:phiSCASCC}
    \Phi_e(\boldsymbol{r}_1,\boldsymbol{r}_2,t)=\sum_{i=1}^{N^{TT}}C_i^{TT}(t)\phi_i^{TT}(\boldsymbol{r}_1,\boldsymbol{r}_2)e^{-iE_i^{TT}t}\\
    +\sum_{j=1}^{N^{PP}}C_j^{PP}(t)\phi_j^{PP}(\boldsymbol{r}_1,\boldsymbol{r}_2)e^{-iE_j^{PP}t}\\
     +\sum_{k=1}^{N^T}\sum_{m=1}^{N^P}C_{km}^{N^TN^P}(t)[\phi_k^T(\boldsymbol{r}_1)\phi_m^P(\boldsymbol{r}_2,t)\\
     + \phi_k^T(\boldsymbol{r}_2)\phi_m^P(\boldsymbol{r}_1,t)]e^{-i(E_k^T+E_m^P)t}
\end{multline}

where the superscripts T and TT (P and PP) describe the states and corresponding energies for one or two electrons on the target (projectile), respectively. In our calculation, the one- and two-electron states centered on the target or projectile are expressed in terms of the Gaussian-type orbitals (GTOs) and of products of these GTOs. The insertion of Eq. (\ref{eqn:phiSCASCC}) into Eq. (\ref{Eq:ham_SCASCC}) results in a system of first-order coupled equations for the scattering amplitudes, which can be written in matrix form as
\begin{equation}
    i\dfrac{\partial}{\partial t}c(t) = S^{-1}(\boldsymbol{b},\boldsymbol{v},t)M(\boldsymbol{b},\boldsymbol{v},t)c(t)
\end{equation}\label{eqn:SCASCC 3}

where c(t) is the column vector of the time-dependent expansion coefficients, S and M are the overlap and coupling matrices, respectively. After solving these equations, the probability ($P_{fi}$) of the transition $i\rightarrow f$ can be obtained. The corresponding cross sections can be calculated from these probabilities as
\begin{equation}
    \sigma_{fi}(v) = 2\int_0^{\infty}bP_{fi}(b,v)db
\end{equation}\label{eqn:SCASCC 4}

\section*{Appendix II}
\subsection*{CTMC simulation}
CTMC simulations are based on calculating a large number of individual particle trajectories, with the initial atomic states selected randomly \cite{R_Abrines_1966,PhysRevA.16.531,TOKESI1994201,K_Tökési_2000}.  
In the model, the Coulomb force acts between the colliding particles, and we calculate the effective charge of the active electron using Slater's rules \cite{PhysRev.36.57}. The randomly selected initial conditions included the impact parameter of the projectile relative to the target atom, as well as the position and velocity vector of the target electron moving in Kepler orbits. 
The initial conditions for individual collisions are chosen at sufficiently large internuclear separations from the collision center, where interactions among the particles are negligible. The classical equations of motion were integrated with respect to time as the independent variable by the standard Runge-Kutta method. Finally, after numerically solving the equations of motion, we computed the total and angular single differential state-selective charge exchange total cross-sections using the following formula:
\begin{equation}
\sigma={{2 \pi b_{max}}\over{T_N}}\sum_j b_j^{(c)},
\label{eq:tothat}
\end{equation}

\begin{equation}
{{d\sigma}\over{d\Omega}}={{2 \pi b_{max}}\over{T_N 
\Delta \Omega}}\sum_j b_j^{(c)},
\label{eq:diffhat}
\end{equation}

\noindent where the associated statistical uncertainty is given by

\begin{equation}
\Delta \sigma= \sigma \left [{{T_N-T_N^{(c)}}\over{T_N-T_N^{(c)}}}
\right ] ^{1/2}.
\label{eq:hiba}
\end{equation}

In Eqs.~\ref{eq:tothat}-~\ref{eq:hiba} $T_N$ 
is the total number of trajectories calculated 
for impact parameters less than $b_{max}$, $T_N^{(c)}$ 
is the number of trajectories that 
satisfy the criteria for capture, and $b_j^{(c)}$
is the actual impact parameter 
for the trajectory corresponding to the capture process under consideration 
in  the emission angle interval 
$\Delta \Omega$ of the electron.

\section*{Appendix III}
\subsection*{Toy model details}
\begin{figure*}
    \centering
\includegraphics[width=\linewidth]{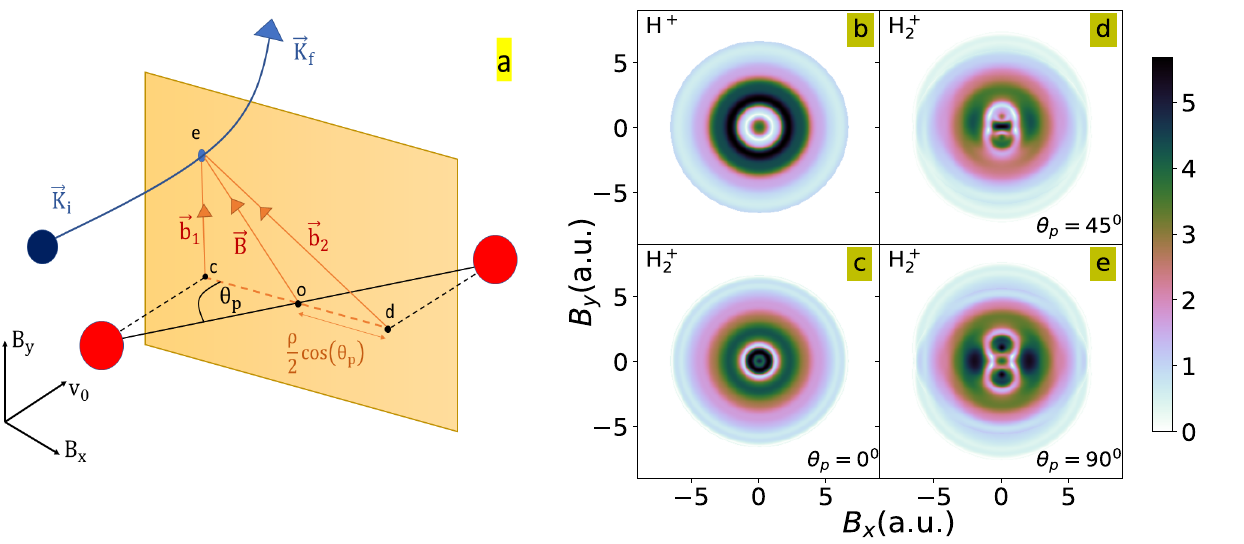}
    \caption{Imaginary Fraunhofer screen evaluated by toy model. In the $\mathrm{H^+}$ projectile rest frame, the origin coincides with the $\mathrm{H^+}$ position. In the  $\mathrm{H_2^+}$ projectile rest frame origin is taken as the center of mass of $\mathrm{H_2^+}$.  Panel (a) depicts the coordinate system in $\mathrm{H_2^+}$ rest frame. Panel (b) presents 2d screen for $\mathrm{H^+}$ projectile. Panels (c), (d), (e) presents 2d screen for the $\mathrm{H_2^+}$ projectile at three different orientations during collision.  Molecular axis was at an angle $\theta_P$ with respect to relative velocity $\mathrm{\boldsymbol{v_0}}$. Color bar represents relative probability amplitude $|\mathcal{\tilde{A}}(\boldsymbol{B})|$.}
    \label{fig:toymodel}
\end{figure*}
For ground state channel, magnetic quantum number remain unchanged ($\Delta m =0$). The simplified expression of equation \ref{eqn:ftp} for this channel can be written as 
\begin{equation}
    f_{i\rightarrow j}^{g.s.}(\Theta,\Phi) = n_c \int_0^\infty \mathcal{\tilde{A}}_{i\rightarrow j}(b)e^{2i\delta_{i\rightarrow j}}J_0(K\Theta b)bdb
    \label{eq:j0}
\end{equation}
where $\mathrm{n_c}$ is a normalizing constant. This equation resembles Fraunhofer diffraction from circular aperture in optics. The imaginary screen $\mathrm{\tilde{A}(b)e^{2i\delta (b)}}$ directly reflects the reaction dynamics \cite{PhysRev.169.84,PhysRevLett.87.123201}. Unique solution could be found by performing an inverse Hankel transformation. If the phase part of the scattering amplitude is known, we only measure the amplitude part through $d\sigma/d\Omega$. In our model, we numerically found a physically acceptable solution by means of inverse curve fitting. The steps are as follows:

Rearranging the screen term as
\begin{equation}
    \mathfrak{A}(b)+i\mathfrak{B}(b) = \mathcal{\tilde{A}}(b)e^{2i\delta (b)}
\end{equation}

where $\mathfrak{A}(b)$ and $\mathfrak{B}(b)$ are real valued functions. We expanded them in Fourier Bessel series
\begin{align}
    \mathfrak{A}(b) &= \sum_{i=0}^n \alpha_i J_0\left(\dfrac{z_i b}{b_{max}-b_{min}}\right) \\
     \mathfrak{B}(b) &= \sum_{i=0}^n \beta_i J_0\left(\dfrac{z_i b}{b_{max}-b_{min}}\right)
\end{align}
where $\alpha_i$ and $\beta_i$ are parameters to be optimized, $\mathrm{z_i}$ is $i^{th}$ zero of Bessel function $\mathrm{J_0}$.

We gave initial guess
\begin{align}
    \mathcal{\tilde{A}}(b) &= \int_0^\infty \sqrt{\dfrac{d\sigma}{d\Omega}} J_0(K\Theta b) K\Theta d(K\Theta) \\
    \delta (b) &= \dfrac{log(b)}{v_0}
\end{align}
    The optimization results are shown in impact parameter dependent probability distribution in fig. (\ref{fig:theory_gs} (a), (b)), and corresponding differential cross sections are shown in fig. (\ref{fig: scattering all} (a), (b)). 

   The two center probability amplitude can be evaluated from one center probability amplitude by following the relation \cite{PhysRevA.40.3673,PhysRevA.40.1302}
    \begin{equation}
        \mathcal{\tilde{A}}_{H_2^+}(\boldsymbol{B}) = \mathcal{\tilde{A}}_{0}(\boldsymbol{b_1})+\mathcal{\tilde{A}}_{0}(\boldsymbol{b_2})e^{i\rho K_\parallel cos(\theta_P)}
        \label{eq:two_center_ab}
    \end{equation}
    
where $\mathcal{\tilde{A}}_0(\boldsymbol{b})$ is the individual center probability amplitude, $\theta_P$ is the angle between $\mathrm{H_2^+}$ molecular axis and relative velocity vector.  The results are shown in fig. (\ref{fig:toymodel}). We can see the screen is always cylindrical symmetric when the projectile is $\mathrm{H^+}$, and for $\mathrm{H_2^+}$, it is cylindrical symmetric when the molecular axis is parallel to relative velocity vector during collision. When the molecular axis is at an angle during collision, cylindrical symmetry breaks due to the interference of two center.

\end{document}